\documentclass[aps,prd,twocolumn,groupedaddress,showpacs]{revtex4}
\usepackage{float,graphicx}

\begin{document}


\title{CMB and Matter Power Spectra of Early $f(R)$ Cosmology in Palatini Formalism}



\author{B.~Li }
\email[Email address: ]{bli@phy.cuhk.edu.hk}
\affiliation{Department of Physics, The Chinese University of Hong
Kong, Hong Kong SAR, China}

\author{M.~-C.~Chu}
\email[Email address: ]{mcchu@phy.cuhk.edu.hk}
\affiliation{Department of Physics, The Chinese University of Hong
Kong, Hong Kong SAR, China}


\date{\today}

\begin{abstract}
We calculate in this article the CMB and matter power spectra of a
class of early $f(R)$ cosmologies, which takes the form of $f(R) =
R + \lambda_1 H_0^2\text{exp}[R/(\lambda_2 H_0^2)]$. Unlike the
late-time $f(R)$ cosmologies such as $f(R) = R + \alpha(-R)^\beta$
($\beta<0$), the deviation from $\Lambda\text{CDM}$ of this model
occurs at a higher redshift (thus the name \emph{Early $f(R)$
Cosmology}), and this important feature leads to rather different
ISW effect and CMB spectrum. The matter power spectrum of this
model is, at the same time, again very sensitive to the chosen
parameters, and LSS observations such as SDSS should constrain the
parameter space stringently. We expect that our results are
applicable at least qualitatively to other models that produce
$f(R)$ modification to GR at earlier times (\emph{e.g.}, redshifts
$\mathcal{O}(10) \lesssim z \lesssim \mathcal{O}(1)$) than when
dark energy begins to dominate -- such models are strongly
disfavored by data on CMB and matter power spectra.
\end{abstract}


\maketitle


\section{INTRODUCTION}

The theory of $f(R)$ gravity has attracted a lot of attention
recently. Theoretically, there are no principles prohibiting the
inclusion of higher order curvature invariants, such as $R^2,
R^{\mu\nu}R_{\mu\nu}$ and
$R^{\alpha\beta\gamma\delta}R_{\alpha\beta\gamma\delta}$ and their
scalar functions, into the Einstein-Hilbert action, as long as the
various constraints from local and cosmological observations are
satisfied. Practically, appropriate choices of these correction
terms could modify the late behavior of the Universe drastically,
thus giving the recently-observed accelerating expansion an
explanation from an alternative theory of gravity. In
\cite{Carroll2005, Easson2005} the authors consider a specific
model in which the correction is a polynomial of $R^2,
R^{\mu\nu}R_{\mu\nu}$ and
$R^{\alpha\beta\gamma\delta}R_{\alpha\beta\gamma\delta}$, and
their analysis shows that there exist late-time accelerating
attractor solutions. Meanwhile, models with $R^2,
R^{\mu\nu}R_{\mu\nu}$ corrections are discussed within the
Palatini approach \cite{Vollick2003, Allemandi2004a,
Allemandi2004b, Allemandi2005}, in which the field equations are
second order, and similar acceleration solutions are found. This
Palatini-$f(R)$ theory of gravitation is then tested using various
cosmological data such as Supernovae (SN), Cosmic Microwave
Background (CMB) shift parameter, baryon oscillation and Big Bang
Nucleosynthesis (BBN), in \cite{Capozziello2006, Amarzguioui2006,
Sotiriou2006}.

Recently, the constraints from CMB and matter power spectra are
obtained for the typical case of $f(R) = R + \alpha(-R)^\beta$ in
\cite{Koivisto2006, Li2006}, which successfully exclude most of
the parameter space, making the model indistinguishable from the
standard $\Lambda\mathrm{CDM}$ in practice. As the corrections to
General Relativity (GR) in this model, which act as an effective
dark energy component, become important only at very late times
(or very low energy densities), these works almost exclude the
possibility of modifying theory of gravity in the $f(R)$-manner at
late times. However, there remains the possibility that $f(R)$
modification to GR occurs in the earlier time. Obviously, the
successful predictions of BBN and primary CMB require that effects
from this modification cannot be significant during these earlier
times. Thus we will concentrate in this work on the case where the
$f(R)$ modification becomes important during some intermediate
times between recombination and present. To be specific, we shall
consider a class of $f(R)$ gravity model with the form
$f(R)=R+\lambda_1 H_0^2\text{exp}(R/\lambda_2 H_0^2)$, the
so-called \emph{exponential $f(R)$ gravity model}, which, as shown
below in Sec.~III~A, well possesses the desirable property of
modifying gravitational theory at intermediate cosmic times
(energy densities). A similar model has been considered in
\cite{Zhang2006} using the metric approach, and an early-time
Integrated Sachs-Wolfe (ISW) effect-Large Scale Structure (LSS)
cross correlation is found there, which is potentially useful for
distinguishing this model from others. Here, however, we shall
analyze it using the Palatini approach and exactly calculate its
CMB and matter power spectra by numerically solving the covariant
perturbation equations given in \cite{Li2006}; comparisons with
the model $f(R)=R+\alpha(-R)^\beta$ and $\Lambda\mathrm{CDM}$
would also be presented and discussed. This work therefore
contributes to filling in the gap in constraining $f(R)$ cosmology
throughout the cosmic history.

This work is organized as the following: in Sec.~II we briefly
introduce the model and list the background and perturbation
evolution equations that are needed in the numerical calculation.
In Sec.~III we incorporate these equations into the public code
CAMB \cite{Lewis2000} and give the numerical results. The
discussion and conclusion are then presented in Sec.~IV.
Throughout this work we will adopt the unit $c=1$ and only
consider the case of a spatially flat (background) Universe
without massive neutrinos.

\section{FIELD EQUATIONS IN $f(R)$ GRAVITY}

In this section we briefly summarize the main ingredients of
$f(R)$ gravity in Palatini formulation and list the general
perturbation equations in this theory.

\subsection{The Modified Einstein Equations}

The starting point of the Palatini-$f(R)$ gravity is the
Einstein-Hilbert action
\begin{eqnarray}
S &=& \int d^4
x\sqrt{-g}\left[\frac{1}{2\kappa}f(R)+\mathcal{L}_{m}\right],
\end{eqnarray}
in which $\kappa = 8\pi G$, $R=g^{ab}R_{ab}(\bar{\Gamma})$ ($a, b
= 0, 1, 2, 3$) with $R_{ab}(\bar{\Gamma})$ being defined as
\begin{eqnarray}
R_{ab} &\equiv& \bar{\Gamma}^{c}_{ab, c} - \bar{\Gamma}^{c}_{ac,
b} + \bar{\Gamma}^{c}_{cd}\bar{\Gamma}^{d}_{ab} -
\bar{\Gamma}^{c}_{ad}\bar{\Gamma}^{d}_{cb}\ .
\end{eqnarray}
Note that the connection $\bar{\Gamma}$ is not the Levi-Civita
connection of the metric $g_{ab}$, which we denote as $\Gamma$;
rather it is treated as an independent field in the Palatini
approach. Correspondingly, the tensor $R_{ab}$ and scalar $R$ are
not the ones calculated from $g_{ab}$ as in GR, which are denoted
by $\mathcal{R}_{ab}$ and $\mathcal{R}$ respectively in this work
($\mathcal{R}=g^{ab}\mathcal{R}_{ab}$). The matter Lagrangian
density $\mathcal{L}_{m}$, however, is assumed to be independent
of $\bar{\Gamma}$, which is the same as in GR.

Varying the action with respect to the metric $g_{ab}$ then leads
to the modified Einstein equations
\begin{eqnarray}
FR_{ab} - \frac{1}{2}g_{ab}f(R) &=& \kappa\mathcal{T}_{ab},
\end{eqnarray}
where $F = F(R) \equiv \partial f(R)/\partial R$ and
$\mathcal{T}_{ab}$ is the energy-momentum tensor. The trace of
Eq.~(3) reads
\begin{eqnarray}
FR - 2f = \kappa\mathcal{T}
\end{eqnarray}
with $\mathcal{T}=\rho - 3p$. This is the so-called structural
equation \cite{Allemandi2004a} which relates $R$ directly to the
energy components in the Universe: given the specific form of
$f(R)$ and thus $F(R)$, $R$ as a function of $\mathcal{T}$ could
be obtained by numerically or analytically solving this equation.

The extremization of the action Eq.~(1) with respect to the
connection field $\bar{\Gamma}$ gives another equation
\begin{eqnarray}
\nabla_{a}[F(R)\sqrt{-g}g^{bc}] &=& 0,
\end{eqnarray}
indicating that the connection $\bar{\Gamma}$ is compatible with
the metric $\gamma_{ab}$ which is conformal to $g_{ab}$:
\begin{eqnarray}
\gamma_{ab} &=& F(R)g_{ab}.
\end{eqnarray}
With Eq.~(6) we could easily obtain the relation between $R_{ab}$
and $\mathcal{R}_{ab}$
\begin{eqnarray}
R_{ab} &=& \mathcal{R}_{ab} +
\frac{3\mathcal{D}_{a}F\mathcal{D}_{b}F}{2F^{2}} -
\frac{\mathcal{D}_{a}\mathcal{D}_{b}F}{F} -
\frac{g_{ab}\mathcal{D}^{c}\mathcal{D}_{c}F}{2F}.
\end{eqnarray}
Note that in the above we use $\mathcal{D}$ and $\nabla$ to denote
the covariant derivative operators compatible with $g_{ab}$ and
$\gamma_{ab}$ respectively.

Since $\mathcal{L}_{m}$ depends only on $g_{ab}$ (and some matter
fields) and the conservation law of the energy-momentum tensor
holds with respect to it, we shall treat this metric as the
physical one. Consequently the difference between $f(R)$ gravity
and GR could be understood as a change of the manner in which the
spacetime curvature and thus the physical Ricci tensor
$\mathcal{R}_{ab}$ responds to the distribution of matter (through
the modified Einstein equations). In order to make this point
explicit, we rewrite Eq.~(3) with the aid of Eq.~(7) as
\begin{eqnarray}
\kappa\mathcal{T}_{ab} &=& F\mathcal{R}_{ab} -
\frac{1}{2}g_{ab}f\nonumber\\
&& + \frac{3}{2F}\mathcal{D}_{a}F\mathcal{D}_{b}F -
\mathcal{D}_{a}\mathcal{D}_{b}F -
\frac{1}{2}g_{ab}\mathcal{D}^{c}\mathcal{D}_{c}F,
\end{eqnarray}
in which $F, f$ are now $F(\mathcal{T}), f(\mathcal{T})$.

\

\subsection{The Perturbation Equations}

The perturbation equations in general theories of $f(R)$ gravity
have been derived in \cite{Koivisto2006b}. However, here we adopt
a different derivation by us \cite{Li2006} which uses the method
of $1+3$ decomposition \cite{Challinor1999, Lewis2000}. For more
details we refer the reader to \cite{Li2006} and here we shall
simply list the results.

The main idea of $1+3$ decomposition is to make space-time splits
of physical quantities with respect to the 4-velocity $u^{a}$ of
an observer. The projection tensor $h_{ab}$ is defined as
$h_{ab}=g_{ab} - u_{a}u_{b}$ which can be used to obtain covariant
tensors perpendicular to $u$. For example, the covariant spatial
derivative $\hat{\mathcal{D}}$ of a tensor field $T^{b\cdot\cdot
c}_{d\cdot\cdot e}$ is defined as
\begin{eqnarray}
\hat{\mathcal{D}}^{a}T^{b\cdot\cdot c}_{d\cdot\cdot e} &\equiv&
h^{a}_{i}h^{b}_{j}\cdot\cdot\ h^{c}_{k}h^{r}_{d}\cdot\cdot\
h^{s}_{e}\mathcal{D}^{i}T^{j\cdot\cdot k}_{r\cdot\cdot s}.
\end{eqnarray}
The energy-momentum tensor and covariant derivative of velocity
could be decomposed respectively as
\begin{eqnarray}
\mathcal{T}_{ab} &=& \pi_{ab} + 2q_{(a}u_{b)} + \rho u_{a}u_{b}
-ph_{ab},\\
\mathcal{D}_{a}u_{b} &=& \sigma_{ab} + \varpi_{ab} +
\frac{1}{3}\theta h_{ab} + u_{a}A_{b}.
\end{eqnarray}
In the above $\pi_{ab}$ is the projected symmetric trace free
(PSTF) anisotropic stress, $q$ the vector heat flux, $p$ the
isotropic pressure, $\sigma_{ab}$ the PSTF shear tensor,
$\varpi_{ab} = \hat{\mathcal{D}}_{[a}u_{b]}$, $\theta =
\mathcal{D}^{c}u_{c} = 3\dot{a}/a$ ($a$ is the cosmic scale
factor) the expansion scalar and $A_{b} = \dot{u}_{b}$ the
acceleration; the overdot denotes time derivative expressed as
$\dot{\phi} = u^{a}\mathcal{D}_{a}\phi$, and the square brackets
mean antisymmetrization and parentheses symmetrization. The
normalization is chosen as $u^{2}=1$.

Decomposing the Riemann tensor and making use of the modified
Einstein equations, we obtain, after linearization, five
constraint equations
\begin{eqnarray}
0 &=& \hat{\mathcal{D}}^{c}(\epsilon^{ab}_{\ \ cd}u^{d}\varpi_{ab});\\
\frac{1}{F}\kappa q_{a} &=&
\frac{3\dot{F}\hat{\mathcal{D}}_{a}F}{2F^{2}} +
\frac{\theta\hat{\mathcal{D}}_{a}F}{3F} -
\frac{\hat{\mathcal{D}}_{a}\dot{F}}{F}\nonumber\\
&& -\frac{2}{3}\hat{\mathcal{D}}_{a}\theta +
\hat{\mathcal{D}}^{b}\sigma_{ab} +
\hat{\mathcal{D}}^{b}\varpi_{ab};\\
\mathcal{B}_{ab} &=& \left[\hat{\mathcal{D}}^{c}\sigma_{d(a}
+ \hat{\mathcal{D}}^{c}\varpi_{d(a}\right]\epsilon_{b)ec}^{\ \ \ \ d}u^{e};\\
\hat{\mathcal{D}}^{b}\mathcal{E}_{ab} &=&
\frac{1}{2F}\kappa\left[\hat{\mathcal{D}}^{b}\pi_{ab} +
\left(\frac{2}{3}\theta + \frac{\dot{F}}{F}\right) q_{a} +
\frac{2}{3}\hat{\mathcal{D}}_{a}\rho\right]\nonumber\\
&& - \frac{1}{2F^{2}}\kappa(\rho + p)\hat{\mathcal{D}}_{a}F;\\
\hat{\mathcal{D}}^{b}\mathcal{B}_{ab} &=&
\frac{1}{2F}\kappa\left[\hat{\mathcal{D}}_{c}q_{d} + (\rho +
p)\varpi_{cd}\right]\epsilon_{ab}^{\ \ cd}u^{b}.
\end{eqnarray}
Here $\epsilon_{abcd}$ is the covariant permutation tensor,
$\mathcal{E}_{ab}$ and $\mathcal{B}_{ab}$ are respectively the
electric and magnetic parts of the Weyl tensor.

Meanwhile, we obtain seven propagation equations:
\begin{eqnarray}
\dot{\rho} + (\rho + p)\theta + \hat{\mathcal{D}}^{a}q_{a} &=& 0;\\
\dot{q_{a}} + \frac{4}{3}\theta q_{a} + (\rho + p)A_{a} -
\hat{\mathcal{D}}_{a}p + \hat{\mathcal{D}}^{b}\pi_{ab} &=& 0;\\
\dot{\theta} + \frac{1}{3}\left[\theta +
\frac{3\dot{F}}{2F}\right]\theta -
\hat{\mathcal{D}}^{a}A_{a}\nonumber\\
-\left[\frac{3\dot{F}^{2}}{2F^{2}} - \frac{3\ddot{F}}{2F} -
\frac{\kappa\rho}{F} - \frac{f}{2F} -
\frac{\hat{\mathcal{D}}^{2}F}{2F}\right] &=& 0;\\
\dot{\sigma}_{ab}+ \frac{2}{3}\left[\theta +
\frac{3\dot{F}}{4F}\right]\sigma_{ab} - \hat{\mathcal{D}}_{\langle
a}A_{b\rangle}\nonumber\\
+ \mathcal{E}_{ab} + \frac{1}{2F}\kappa\pi_{ab} +
\frac{1}{2F}\hat{\mathcal{D}}_{\langle
a}\hat{\mathcal{D}}_{b\rangle}F &=& 0;\\
\dot{\varpi} + \frac{2}{3}\theta\varpi -
\hat{\mathcal{D}}_{[a}A_{b]}
&=& 0;\\
\frac{1}{2F}\kappa\left[\dot{\pi}_{ab} + \left(\frac{1}{3}\theta -
\frac{3\dot{F}}{2F}\right)\pi_{ab}\right]\nonumber\\ -
\frac{1}{2F}\kappa\left[(\rho + p)\sigma_{ab} \ +
\hat{\mathcal{D}}_{\langle a}q_{b\rangle}\right]\nonumber\\
-\left[\dot{\mathcal{E}}_{ab} + \left(\theta +
\frac{\dot{F}}{2F}\right)\mathcal{E}_{ab} -
\hat{\mathcal{D}}^{c}\mathcal{B}_{d(a}\epsilon_{b)ec}^{\ \ \ \ d}u^{e}\right] &=& 0;\\
\dot{\mathcal{B}}_{ab} + \left(\theta +
\frac{\dot{F}}{2F}\right)\mathcal{B}_{ab} +
\hat{\mathcal{D}}^{c}\mathcal{E}_{d(a}
\epsilon_{b)ec}^{\ \ \ \ d}u^{e} \nonumber\\
+ \frac{1}{2F}\kappa\hat{\mathcal{D}}^{c}
\mathcal{\pi}_{d(a}\epsilon_{b)ec}^{\ \ \ \ d}u^{e} &=& 0,
\end{eqnarray}
where the angle bracket means taking the trace free part of a
quantity.

Besides the above equations, it would also be useful to express
the projected Ricci scalar $\hat{\mathcal{R}}$ in the
hypersurfaces orthogonal to $u^{a}$ as \cite{Li2006}
\begin{eqnarray}
\hat{\mathcal{R}} &\doteq& \frac{\kappa(\rho + 3p)-f}{F} -
\frac{2}{3}\left[\theta + \frac{3\dot{F}}{2F}\right]^{2} -
\frac{2\hat{\mathcal{D}}^{2}F}{F}.
\end{eqnarray}
The spatial derivative of the projected Ricci scalar, $\eta_{a}
\equiv \frac{1}{2}a\hat{\mathcal{D}}_{a}\hat{\mathcal{R}}$, is
then given as
\begin{eqnarray}
\eta_{a} &=& \frac{a}{2F}\kappa(\hat{\mathcal{D}}_{a}\rho +
3\hat{\mathcal{D}}_{a}p) - \frac{a}{F}\left[\frac{3}{2F}\dot{F} +
\theta\right]\hat{\mathcal{D}}_{a}\dot{F}\nonumber\\
&& - \frac{a}{2F}\hat{\mathcal{D}}_{a}f -
\frac{a}{F}\hat{\mathcal{D}}_{a}(\hat{\mathcal{D}}^{2}F) -
\frac{2a}{3}\left[\frac{3}{2F}\dot{F} +
\theta\right]\hat{\mathcal{D}_{a}}\theta\nonumber\\
&& + \frac{a}{3F}\left[\frac{3}{2F}\dot{F} +
\theta\right]\left[\frac{3}{2F}\dot{F} -
\theta\right]\hat{\mathcal{D}}_{a}F,
\end{eqnarray}
and its propagation equation
\begin{eqnarray}
\dot{\eta}_{a} + \frac{2\theta}{3}\eta_{a} &=&
\frac{a}{2F}\left[\frac{3}{F}\dot{F} -
\frac{2}{3}\theta\right]\hat{\mathcal{D}}_{a}\hat{\mathcal{D}}^{2}F
-
\frac{a}{F}\kappa\hat{\mathcal{D}}_{a}\hat{\mathcal{D}}^{c}q_{c}\nonumber\\
&& -
\frac{a}{F}\hat{\mathcal{D}}_{a}(\hat{\mathcal{D}}^{2}F)^{\cdot} -
\left[\frac{\dot{F}}{F} +
\frac{2\theta}{3}\right]a\hat{\mathcal{D}}_{a}\hat{\mathcal{D}}^{c}A_{c}.\
\ \ \ \
\end{eqnarray}

Since we are considering a spatially flat Universe, the spatial
curvature must vanish for large scales which means that
$\hat{\mathcal{R}}=0$. Thus from Eq.~(24) we obtain
\begin{eqnarray}
\left[\frac{1}{3}\theta + \frac{\dot{F}}{2F}\right] ^{2} &=&
\frac{1}{6F}\left[\kappa(\rho+3p)-f\right].
\end{eqnarray}
This is just the modified (first) Friedmann equation of $f(R)$
gravity, and the other modified background equations (the second
Friedmann equation and the energy-conservation equation) could be
obtained by taking the zero-order parts of Eqs.~(17, 19). It is
easy to check that when $f(R) = R$, we have $F = 1$ and these
equations reduce to those in GR.

Recall that we have had $f, F$ and $R$ as functions of
$\mathcal{T}$; it is then straightforward to calculate $\dot{F},
\ddot{F}, \hat{\mathcal{D}}_{a}F, \hat{\mathcal{D}}_{a}\dot{F}$
\emph{etc.}~as functions of $\dot{\mathcal{T}} \doteq - (\rho_{b}
+ \rho_{c})\theta$ and $\hat{\mathcal{D}}_{a}\mathcal{T} =
(1-3c_{s}^{2})\hat{\mathcal{D}}_{a}\rho_{b} +
\hat{\mathcal{D}}_{a}\rho_{c}$, in which $\rho_{b(c)}$ is the
energy density of baryons (cold dark matter) and $c_{s}$ is the
baryon sound speed. Notice that we choose to neglect the small
baryon pressure except in the terms where its spatial derivative
is involved, as they might be significant at small scales. The
above equations could then be numerically propagated given the
initial conditions, telling us information about the evolutions of
small density perturbations and the CMB anisotropies in theories
of $f(R)$ gravity. These results will be given in the next
section.

\section{NUMERICAL RESULTS}

Now let us consider the model $f(R) = R + \lambda_1
H_0^2\mathrm{exp}(R/\lambda_2 H_0^2)$. There are two positive
dimensionless parameters in this model, $\lambda_1$ and
$\lambda_2$, of which, roughly speaking, $\lambda_1$ controls the
overall significance of the correction to GR while $\lambda_2$
determines the time at which the correction becomes important: at
early times when $|R| \gg \lambda_2 H_0^2$, the correction
$\lambda_1 H_0^2\mathrm{exp}(R/\lambda_2 H_0^2) \rightarrow 0$ and
at late times when $|R| \ll \lambda_2 H_0^2$ it tends to a
constant $\lambda_1 H_{0}^2$, thus retrieving the
$\Lambda\mathrm{CDM}$ model; deviation from $\Lambda\mathrm{CDM}$
becomes significant when $|R| \sim \lambda_2 H_0^2$. Solar system
constraint on $\lambda_2$ is estimated in \cite{Zhang2006} to be
$\lambda_2 \ll 10^{6}$, which is also applicable here. $H_{0}$ is
a constant with dimension $\mathrm{Mpc}^{-1}$ which helps to make
$\lambda_1, \lambda_2$ dimensionless and we will take it to be the
currently preferred value of the Hubble constant, \emph{i.e.},
$H_{0} = 72\ \mathrm{km\ s^{-1}\ Mpc^{-1}}$, for later
convenience.

Just like in the case of $f(R) = R + \alpha(-R)^\beta$, it could
be easily checked that the parameters $\lambda_1, \lambda_2$ and
$\Omega_m$ (the dimensionless density parameter of baryon $+$ cold
dark matter) are not independent. We thus can choose $\lambda_2,
\Omega_m$ as the independent parameters and numerically calculate
$\lambda_1$ from Eqs.~(4, 27). For purpose of illustration we will
fix $\Omega_m = 0.3$ in this work.

\subsection{The Background Evolution}

The background evolution of this model with different values of
$\lambda_2$ is shown in Fig.~1. The results are rather similar to
those obtained with the metric approach \cite{Zhang2006}. As
expected, the deviation from $\Lambda\mathrm{CDM}$ is maximized at
some intermediate ages and reduced towards both earlier and later
times. At $z \lesssim 2$, the deviations are less than $\sim
0.2\%$ and $\sim 0.5\%$ for $\lambda_2 = 1000$ and $500$; such
small deviations are difficult to be identified by the
measurements of $H(z)$ such as SNe observations. However, as shown
below, their effects on the CMB and matter power spectra could be
rather large and $\lambda_2 = 1000$ could be safely excluded by
current observation on matter power spectrum alone.

The evolution of $F(R) - 1$ provides another measurement of the
deviation from $\Lambda\mathrm{CDM}$ because in the latter $F$ is
simply 1. It is clear that $F$ increases at some earlier times and
finally tends to another constant larger than 1. Since, from
Eq.~(27), the effective gravitational constant is $G/F$, the model
becomes effectively $\Lambda\mathrm{CDM}$ with a smaller Newton's
constant at its later evolutionary stage. However, we expect this
effect to be small \cite{Zhang2006} for large enough $\lambda_2$.

\begin{figure}[]
\includegraphics[scale=0.4]{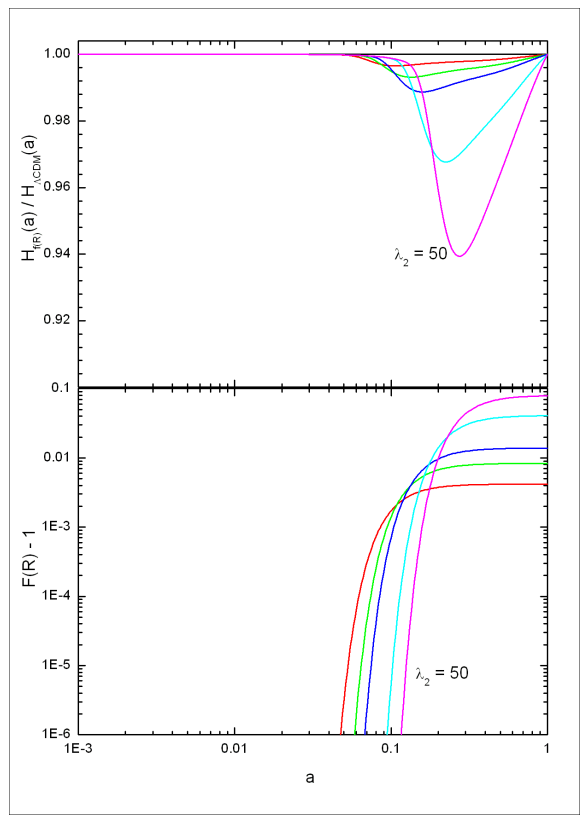}
\caption{(Color online) The background evolutions. Upper panel:
the ratio between the expansion rates in this exponential $f(R)$
cosmology and in $\Lambda\mathrm{CDM}$ as a function of the scale
factor $a$. Lower panel: $F(R)-1$ as a function of $a$. The black,
red, green, blue, cyan and magenta curves represent cases of
$\lambda_2 = \infty, 1000, 500, 300, 100, 50$ respectively.}
\end{figure}

\subsection{The TT CMB Spectrum}

\begin{figure}[]
\includegraphics[scale=0.4]{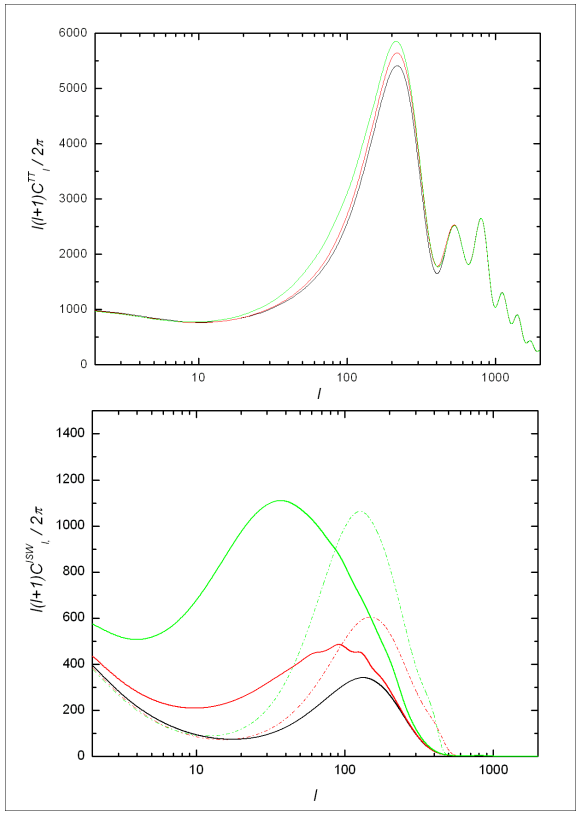}
\caption{(Color online) Upper panel: The theoretical CMB spectrum
of the present model. From bottom to top: $\lambda_2 = \infty,
1000, 500$. Lower panel: Contributions from the ISW effect. Solid
curves are the results of the $f(R) = R + \lambda_1
H_0^2\mathrm{exp}(R/\lambda_2 H_0^2)$ model, from bottom to top:
$\lambda_2 = \infty, 1000, 500$; Dashed curves are those of the
$f(R) = R + \alpha(-R)^\beta$ model for comparison, from bottom to
top: $\beta = 0, -0.05, -0.1$. In both cases we use $\Omega_m =
0.3$.}
\end{figure}

We have calculated the CMB spectrum of the present model using
CAMB, with adiabatic initial conditions because the corrections to
GR are expected not to influence initial conditions. The results
are shown in Fig.~2, where we have plotted the CMB spectra for
$\lambda_2 = \infty, 1000$ and 500. Interestingly we see that the
deviation from $\Lambda\mathrm{CDM}$ ($\lambda_2 = \infty$) occurs
roughly in the range $l\in (10, 500)$, while for even smaller
$l$'s it vanishes. This is in contrast to the $f(R) = R +
\alpha(-R)^\beta$ model, in which drastic changes happen at $l
\lesssim 300$ \cite{Li2006}. The contributions from ISW effect
alone are also shown in Fig.~2, both for the $f(R) = R + \lambda_1
H_0^2\mathrm{exp}(R/\lambda_2 H_0^2)$ and for the $f(R) = R +
\alpha(-R)^\beta$ models. Obviously these agree qualitatively with
the behaviors of the modified CMB spectra in the two models, since
at low $l$'s a significant late-time contribution comes from the
ISW effect.

The ISW effect involves the integral of the time-variation of the
potential weighted by the spherical Bessel functions for each $l$
\begin{eqnarray}
I_{l}^{ISW}(k) &=&
2\int_{0}^{\eta_{0}}e^{-\tau}\phi'_{k}j_{l}[k(\eta_{0}-\eta)]d\eta,
\end{eqnarray}
in which $\tau$ is the optical depth, $\eta$ is the conformal
time, a prime denotes derivative with respect to $\eta$ and
subscript $'$0$'$ means the current value; $\phi_{k}$ is Weyl
tensor variable defined as the coefficient of harmonic expanding
$\mathcal{E}_{ab}$ in terms of $Q^{k}_{ab}$ \cite{Lewis2000}
\begin{eqnarray}
\mathcal{E}_{ab} &=&
-\sum_{k}\frac{k^{2}}{a^{2}}\phi_{k}Q^{k}_{ab},
\end{eqnarray}
where $Q^{k}_{ab} = \frac{a^{2}}{k^{2}}\hat{\mathcal{D}}_{\langle
a}\hat{\mathcal{D}}_{b\rangle}Q^{k}$ and $Q^{k}$ is the
eigenfunction of $\hat{\mathcal{D}}^{2}$ with eigenvalue
$k^{2}/a^{2}$.

\begin{figure}[]
\includegraphics[scale=0.4]{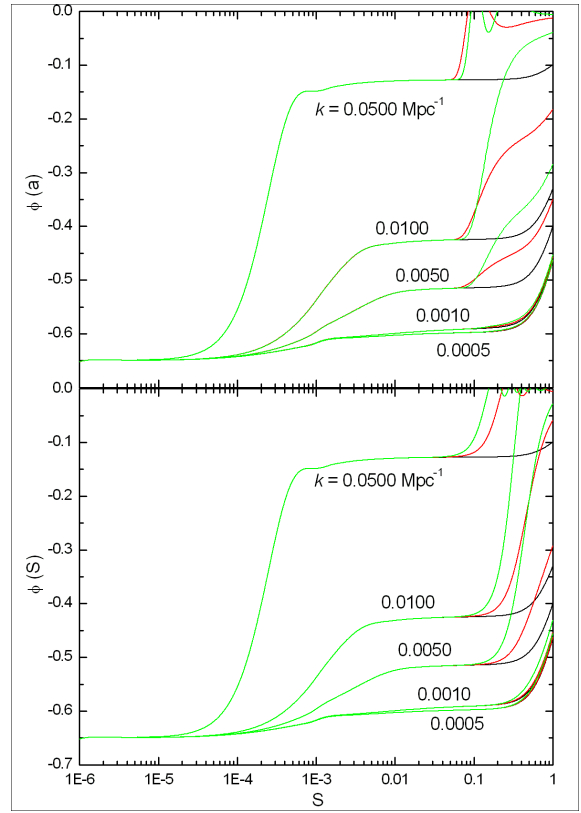}
\caption{(Color online) Upper panel: The potential $\phi_{k}$ as a
function of the scale factor for the different $k$-modes as
indicated besides the curves for the $f(R) = R + \lambda_1
H_0^2\mathrm{exp}(R/\lambda_2 H_0^2)$ model. The black, red and
green curves represent cases of $\lambda_2 = \infty, 1000$ and 500
respectively. Lower panel: same as above, but for the $f(R) = R +
\alpha(-R)^\beta$ model, and the black, red and green curves here
represent cases of $\beta = 0, -0.01$ and -0.05. In both cases we
use $\Omega_m = 0.3$. }
\end{figure}

\begin{figure}[]
\includegraphics[scale=0.5]{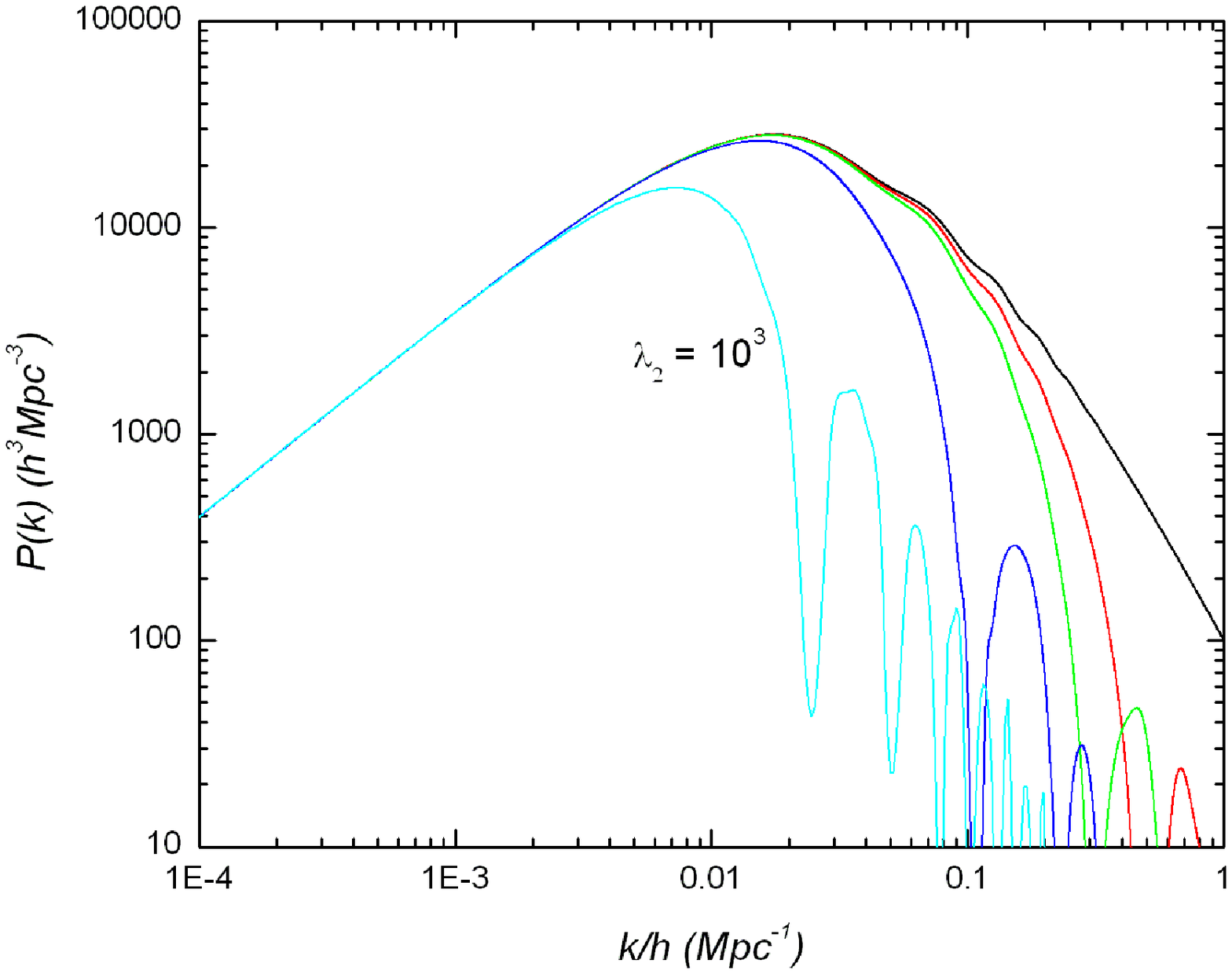}
\caption{(Color online) The matter power spectrum in the present
model. The black, red, green, blue and cyan curves represent the
cases of $\lambda_2 = \infty, 10^{5}, 5\times10^{4}, 10^{4}$ and
1000 respectively.}
\end{figure}

The evolutions of the potential $\phi_{k}$ for various $k$'s are
plotted in Fig.~3, and these provide a qualitative explanation for
Fig.~2. For the ISW effect, we know that the photons typically
travel through many peaks and valleys of the perturbation, and
this leads to cancellations of the effects. Mathematically this
arises from the oscillatory property of the spherical Bessel
function $j_{l}(x)$: $\int dx j_{l}(x) =
\sqrt{\pi}\Gamma[\frac{1}{2}(l+1)]/2\Gamma[\frac{1}{2}(l+2)]
\approx (\pi/2l)^{1/2}$ \cite{Hu2002}. Thus if $\phi'_{k}$ is a
constant, then $I^{ISW}_{l}(k)$ would roughly scale as $l^{-1/2}$
for a specified $k$. If $\phi'_{k}$ is not a constant, on the
other hand, it must be weighted by $j_l(x)$ at different times in
the integration, and, as $j_l(x)$ strongly peaks at $x \sim l$, it
is weighted more at $x \sim l$ for an $l$. To be more explicit,
for the scales we are interested in the ISW integral takes the
following form \cite{Hu1995}
\begin{eqnarray}
\int^{\eta_{0}}e^{-\tau}\phi'_{k}j_{l}[k(\eta_{0}-\eta)]d\eta
\simeq e^{-\tau}\phi'_{k}(\eta_{k})\frac{\sqrt{\pi}}{2k}
\frac{\Gamma[\frac{1}{2}(l+1)]}{\Gamma[\frac{1}{2}(l+2)]},\nonumber
\end{eqnarray}
in which $\eta_{k} = \eta_{0} - (l + 1/2)/k$ is the position of
the main peak of $j_{l}$.

From Fig.~3 it is clear that $\phi_{k}$ decays rapidly earlier in
the $f(R) = R + \lambda_1 H_0^2\mathrm{exp}(R/\lambda_2 H_0^2)$
model than in the $f(R) = R + \alpha(-R)^\beta$ model (for such a
comparison we could choose $\lambda_2$ and $\beta$ such that their
endpoints of the evolution paths of $\phi_{k}$ are nearly the same
in Fig.~3, \emph{e.g.}, $\lambda_2 = 500$ and $\beta = -0.01$
there). On the other hand, at very late times the evolution of
$\phi_{k}$ in the exponential $f(R)$ gravity is nearly the same as
in $\Lambda\mathrm{CDM}$. Now consider the contribution from a
specified $k$-mode: if $l$ is very large then $\eta_{k}$ will be
so small that $\phi'_{k}(\eta_{k}) \simeq 0$; thus the
contribution to this $l$ is negligible. If $l$ is very small, then
the contribution to this $l$ will be similar to that in
$\Lambda\mathrm{CDM}$ as $\phi'_{k}(\eta_{k}) \simeq
\phi'_{k\Lambda\mathrm{CDM}}(\eta_{k})$. For the intermediate
$l$'s, since the larger $l$ is the smaller $\eta_{k}$ is, some
sufficiently larger $l$'s would receive significant contribution
from the ISW effect if the massive decay of the potential occurs
earlier (at a time closer to $\eta_{k}$). Qualitatively this is
why in Fig.~2 the ISW effect for the $f(R) = R + \lambda_1
H_0^2\mathrm{exp}(R/\lambda_2 H_0^2)$ model extends to larger
$l$'s than that of the $f(R) = R + \alpha(-R)^\beta$ model and the
differences from $\Lambda\text{CDM}$ are reduced dramatically
towards very-small $l$'s in the case of exponential $f(R)$
gravity.

As for the CMB polarization spectra of the model, since they are
not as sensitive to the parameters as the temperature spectrum, we
shall not discuss them here.

\subsection{The Matter Power Spectrum}

\begin{figure}[]
\includegraphics[scale=0.4]{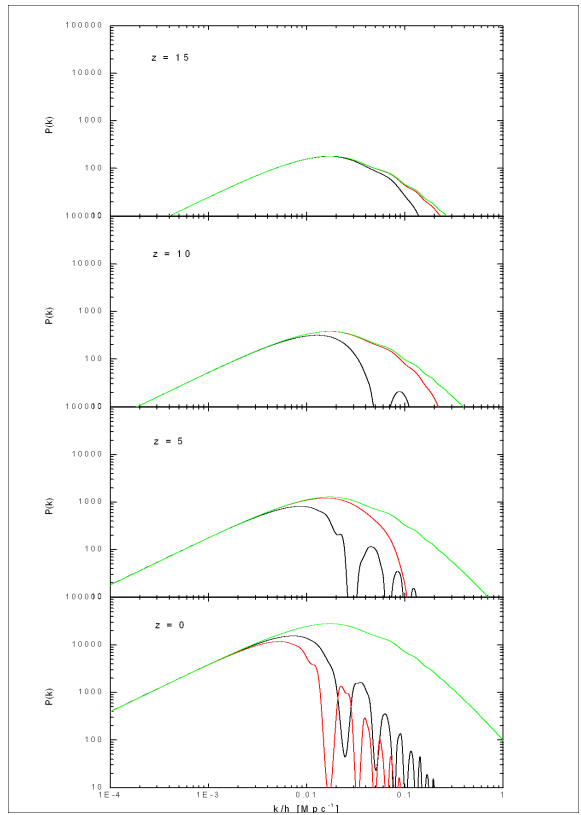}
\caption{(Color online) The matter power spectra at different
redshifts. From bottom to top: $z=0$, $z=5$, $z=10$ and $z=15$.
The black, red, and green curves represent the cases of $f(R)=R +
\lambda_1 H_0^2\mathrm{exp}(R/\lambda_2 H_0^2)$ ($\lambda_2 =
1000$), $f(R) = R + \alpha(-R)^\beta$ ($\beta = -0.01$) and
$\Lambda\mathrm{CDM}$ respectively.}
\end{figure}

As discussed extensively in \cite{Koivisto2006, Koivisto2006b},
the response of this modified gravity to the spatial variations of
matter distribution could be very sensitive (this could also be
seen from Eq.~(24) above), and at small enough scales this feature
would significantly affect the growth of density perturbations. To
be explicit, at large enough $k$'s, the growth equation for the
comoving energy density fluctuations $\delta_m$ could be written
as \cite{Koivisto2006}
\begin{eqnarray}
\frac{d^{2}\delta_m}{dx^{2}} &\doteq& -
\frac{k^{2}}{H^{2}}\frac{F'}{3F(2FH+F')}\delta_m,
\end{eqnarray}
in which $x \equiv \log a$ and $F'/3F(2FH+F')=c_s^2$ acts as an
effective sound speed squared. For the present exponential $f(R)$
theory we have seen from Fig.~1 that $F'>0$, so that $c_s^2>0$ and
the small scale density fluctuations will oscillate instead of
growing according to Eq.~(30). On the other hand, if $c_s^{2}<0$,
then these small scale fluctuations will become unstable and blow
up dramatically \cite{Sandvik2004}; this does happen in the $f(R)
= R + \alpha(-R)^\beta$ model if $\beta$ is positive
\cite{Li2006}, but not for our choices of parameters in the
$f(R)=R + \lambda_1 H_0^2\mathrm{exp}(R/\lambda_2 H_0^2)$ model.

We have also calculated the matter power spectrum in the present
model for different choices of $\lambda_2$ with the modified CAMB
code and the results are shown in Fig.~4. As expected, the matter
spectrum is more sensitive to the deviation from GR than the CMB
spectrum, because it relates directly to the density
perturbations. From this figure we see that $\lambda_2 \lesssim
5\times10^{4}$ will hopefully be excluded when tested against the
measurements on matter power spectrum from Sloan Digital Sky
Survey (SDSS) \cite{Tegmark2004}, which is more stringent than the
possible CMB constraints.

It will also be interesting to consider when the deviations of the
matter power spectrum from the $\Lambda\mathrm{CDM}$ case occur in
the $f(R)$ gravity theories. Since $c_s^2=0$ in the strict
$\Lambda\mathrm{CDM}$ model, the deviation will appear when $F'$
is significantly nonzero, and it thus appears earlier if $F$
begins to evolve earlier. Fig.~5 confirms this expectation: it is
apparent that for the $f(R)=R + \lambda_1
H_0^2\mathrm{exp}(R/\lambda_2 H_0^2)$ model the deviation from the
$\Lambda\mathrm{CDM}$ matter power spectrum has almost shaped up
before $z\simeq5$ while for the $f(R) = R + \alpha(-R)^\beta$
model this happens much later.

\section{Discussion and Conclusions}

To summarize, we have investigated the background evolution, CMB
and matter power spectra of a class of early $f(R)$ gravity model,
which takes the form of $f(R)=R + \lambda_1
H_0^2\mathrm{exp}(R/\lambda_2 H_0^2)$. Such a model has the
interesting feature that the deviation from $\Lambda\mathrm{CDM}$
happens only at some intermediate time, and this is why its CMB
and matter power spectra are rather different from those in either
the $\Lambda\mathrm{CDM}$ case or some late-time $f(R)$ gravity
theories, such as $f(R) = R + \alpha(-R)^\beta$. For the CMB
spectrum, the earlier massive decay of the potential extends the
ISW effect to larger $l$'s, leaving the ISW at very low $l$'s
indistinguishable from that in $\Lambda\mathrm{CDM}$. Consequently
the deviation from the latter is most significant at intermediate
$l$'s ($l \in (10, 400\sim500)$), as shown in Fig.~2. For the
matter power spectrum, the modifications to GR lead to the
appearance of effective pressure fluctuations, which restricts the
growths of small-scale density perturbations and leads to
oscillations in small scales of the spectrum. It is interesting
that, the deviation from $\Lambda\mathrm{CDM}$ matter power
spectrum develops rather early in the early $f(R)$ gravity model
because $F$ here begins to evolve earlier.

Since the matter power spectrum directly shows the density
perturbations while the CMB spectrum does not, the former is more
sensitive to corrections to GR in $f(R)$ gravity, as shown in
\cite{Li2006}. The constraints from local (solar system)
measurements plus the SDSS data might be able to restrict the
parameter $\lambda_2$ into a rather narrow range, roughly
$5\sim10\times10^4\lesssim\lambda_2 \ll 10^{6}$, but an exact
constraint obtainable through carefully searching the parameter
space with, say, the Markov Chain Monte Carlo method is beyond the
scope of the work here.

The present work therefore contributes to filling in the gap in
constraining $f(R)$ corrections to GR at some higher redshifts
than when the late-time accelerating expansion of our Universe
begins; it shows that such early $f(R)$ cosmologies are strongly
disfavored by current data on CMB and matter power spectra, just
as their late-time correspondences such as the model $f(R) =  R +
\alpha(-R)^\beta$.

\begin{acknowledgments}
The work described in this paper was partially supported by a
grant from the Research Grants Council of the Hong Kong Special
Administrative Region, China (Project No.~400803). We are grateful
to Professor Tomi Koivisto for his helpful discussion.
\end{acknowledgments}

\appendix

\newcommand{\noopsort}[1]{} \newcommand{\printfirst}[2]{#1}
  \newcommand{\singleletter}[1]{#1} \newcommand{\switchargs}[2]{#2#1}

\end{document}